\documentclass[showpacs,aps,prl,twocolumn,epsfig,superscriptaddress]{revtex4}
\usepackage{amssymb}
\usepackage{amsmath}
\usepackage{graphicx}
\usepackage{epsfig}

\DeclareMathOperator{\Real}{Re}

\begin{document}

\title{Strong electron-phonon interaction in multiband superconductors.}
\author{ O.V. Dolgov}
\affiliation{Max-Planck-Institut f\"{u}r Festk\"{o}rperphysik, Heisenbergstr.1, 70569
Stuttgart, Germany}
\author{A.A. Golubov}
\affiliation{Faculty of Science and Technology and MESA+ Institute for Nanotechnology,
University of Twente, 7500 AE Enschede, The Netherlands}

\begin{abstract}
We discuss the effects of anisotropy on superconducting critical temperature
and order parameter in a strongly coupled regime. The multiband
representation is used as a model for anisotropy. We show that strong
coupling effects in multiband superconductors lead to pair-breaking due to
interband coupling because soft phonon modes play the same role as usual
impurities. This effect makes the order parameters in different bands equal
to each other and limits the upper bound on critical temperature.
\end{abstract}

\date{\today }
\maketitle

\section{Introduction}

Effects of anisotropy on superconducting critical temperature and energy gap
become of primary importance by approaching the strong-coupling regime when
transition temperature $T_{c}$ becomes of the order or even larger than the
characteristic energy $\Omega $ of a boson modes which mediate
superconductivity. This issue received little attention up to now. In the
weak-coupling limit, the effects of anisotropy were investigated shortly
after the Bardeen, Cooper, Schrieffer (BCS) theory ( see, e.g., Ref. \cite%
{pokr}, and for multiband systems Refs. \cite{mosk},\cite{smw}). Following
the paper by Markovitz and Kadanoff \cite{MK}, different authors (references
can be found in the review \cite{AllenMitr}) introduced the so-called
separable interaction
\begin{equation}
V_{\mathbf{kk}^{\prime }}=(1+a_{\mathbf{k}})V(1+a_{\mathbf{k}^{\prime }}),  \label{sep}
\end{equation}%
where $a_{\mathbf{k}}$ is an anisotropy parameter, the Fermi surface
averaging $\left\langle a_{\mathbf{k}}\right\rangle $ being equal to zero.
The result is the enhancement of the effective coupling constant

\begin{equation*}
\lambda _{eff}=\left\langle N(0)V_{\mathbf{kk}^{\prime }}\right\rangle
=N(0)V(1+\left\langle a_{\mathbf{k}}^{2}\right\rangle )>N(0)V
\end{equation*}%
and corresponding rising of the of $T_{c}$ according to the standard BCS
expression%
\begin{equation}
T_{c}=1.14\theta _{D}\exp (-1/\lambda _{eff}),  \label{tcBCS}
\end{equation}%
where $\theta _{D}$ is the phonon cut-off.

For multiband clean systems in the weak-coupling limit the effective
coupling constant in Eq.\ref{tcBCS} is determined by the \textit{maximum
eigenvalue }of the matrix $\lambda _{\alpha \beta }$, where $\alpha ,\beta $
are band indices. \textit{Intraband} impurity scattering does not affect
superconducting properties (Anderson's theorem), while the \textit{interband}
one averages out the order parameters $\Delta _{\alpha }$ and $\lambda _{eff}
$ (and $T_{c}$) corresponds to the average value

\begin{equation}
\left\langle \lambda \right\rangle =\frac{\sum_{\alpha \beta }N_{\alpha
}(0)\lambda _{\alpha \beta }}{\sum_{\alpha }N_{\alpha }(0)},  \label{av}
\end{equation}
(see e.g. Refs.\cite{mazin93}, \cite{gol}). For positively defined matrix $
\lambda _{\alpha \beta }$ the maximum eigenvalue is bigger than $%
\left\langle \lambda \right\rangle $, and we have the enhancement of $T_{c}$
for multiband systems in comparison with the averaged value independently on
the sign of the nondiagonal matrix elements which determine the anisotropic
contribution \cite{Rainer}.

Recent theoretical studies of superconductivity in the two-band
superconductor $MgB_{2\text{ }}$ \cite{MgB2} and calculations of covalent
metals as the hypothetical hexagonal LiB and boron-doped diamond renewed the
interest to the problem of an upper bound on superconducting critical
temperature in strongly coupled anisotropic systems. Some estimates provide
values of $\lambda$ in anisotropic superconductors as large as 4 (Ref.\cite%
{moussa}) or even 25 (Ref.\cite{Pickett}).

Let us first remind the result for the strong coupling approach to isotropic
systems. For the case $\Omega <<2\pi T_{c}$ (which can occur for large $%
\lambda $) real phonons give the pairbreaking contributions to the
superconducting pairing as well as to the quasiparticle renormalization. The
largest terms corresponding to pair-breaking and quasiparticle damping (see
Appendix A) cancel each other (\cite{kmmProblema,VIK}) and as the result one
arrives the following strong coupling expression (see Ref.\cite{AllenDynes})
\begin{equation}
T_{c}=const\sqrt{\lambda \Omega ^{2}},  \label{strong}
\end{equation}%
where in the simplest approximation $const=(2\pi )^{-1}\simeq 0.15$
(numerical calculations give $0.1827$). There are interpolation expressions
connecting strong- and weak-coupling limits (see reviews \cite%
{AllenMitr,DM,carbotte}).

The authors of Ref.\cite{moussa} have imposed two possible upper bounds on a
maximal critical temperature of multiband superconductors: the lower one is
determined by the averaged coupling constant (\ref{av}), while the upper one
is governed by the maximal (positive) eigenvalue of the matrix for the first
momentum of the Eliashberg functions $\alpha _{\alpha \beta }^{2}(\omega
)F_{\alpha \beta }(\omega )$
\begin{equation}
\left[ \lambda \Omega ^{2}\right] _{\alpha \beta }=M_{\alpha \beta
}(1)=2\int_{0}^{\infty }d\omega \omega \alpha _{\alpha \beta }^{2}(\omega
)F_{\alpha \beta }(\omega )  \label{M1}
\end{equation}%
( for the Einstein spectrum this value is equal to $\lambda _{\alpha \beta
}\Omega ^{2}$).

The purpose of this work is to analyze selfconsistently the effects of
anisotropy on the upper bound on $T_c$. We show that the low frequency
phonons play a role similar to \textit{\ intraband} and \textit{interband}
static impurities. The latter can lead to the suppression of the anisotropy
and as a result the upper bound on $T_c$ is determined by the averaged
coupling constant. We consider in a more detail the applications to the
multiband systems.

\section{General description of multiband systems}

\bigskip The gap functions $\Delta _{\alpha }\left( \omega _{n}\right) $ can
be calculated within an extension of the Eliashberg formalism to two bands
\begin{equation}
\Delta _{\alpha }\left( \omega _{n}\right) Z_{\alpha }\left( \omega
_{n}\right) =\pi T\sum_{\beta }\sum_{\left\vert \omega _{m}\right\vert \leq
\omega _{c}}\frac{\left( \lambda _{\alpha \beta }-\mu _{\alpha \beta }^{\ast
}\right) \Delta _{\beta }\left( \omega _{m}\right) }{\sqrt{\omega
_{m}^{2}+\Delta _{\beta }^{2}\left( \omega _{m}\right) }},
\label{EliashbergG}
\end{equation}%
\begin{equation}
Z_{\alpha }\left( \omega _{n}\right) =1+\frac{\pi T}{\omega _{n}}\sum_{\beta
}\sum_{\omega _{m}}\lambda _{\alpha \beta }\frac{\omega _{m}}{\sqrt{\omega
_{m}^{2}+\Delta _{\beta }^{2}\left( \omega _{m}\right) }},
\label{EliashbergZ}
\end{equation}%
where
\begin{equation*}
\lambda _{\alpha \beta } =2\int_{0}^{\infty }{\omega \alpha _{\alpha \beta
}^{2}(\omega)F_{\alpha \beta }\left( \omega \right) d\omega }/[{\omega
^{2}+\left( \omega _{m}-\omega _{n}\right) ^{2}}],
\end{equation*}%
$Z_{\alpha }\left( \omega _{n}\right) $ are the Migdal renormalization
functions and $\omega _{n}=\pi T(2n-1)$ and the standard Eliashberg
functions define the superconducting properties and thermodynamical
properties are
\begin{eqnarray}
\alpha _{\alpha \beta }^{2}(\omega )F_{\alpha \beta }(\omega ) &=&\frac{1}{%
N_{\alpha }(0)}\sum_{\mathbf{k,k}^{\prime },\nu }\left\vert g_{\mathbf{k,k}%
^{\prime }}^{\alpha \beta ,\nu }\right\vert ^{2}\delta (\varepsilon _{%
\mathbf{k}}^{\alpha })\delta (\varepsilon _{\mathbf{k^{\prime }}}^{\beta })
\notag \\
&&\times \delta (\omega -\omega _{\mathbf{k-k^{\prime }}}^{\nu }),
\label{eliashSC}
\end{eqnarray}%
where $\alpha ,\beta =\{1,2,...\},$ $N_{\alpha }(0)$ is the partial density
of states per spin at the Fermi energy, $g_{\mathbf{k,k}^{\prime }}^{\alpha
\beta }$ is the electron-phonon interaction (EPI) matrix element. Defining $%
\lambda _{\alpha \beta }=2\int \omega ^{-1}\alpha _{\alpha \beta
}^{2}(\omega )F_{\alpha \beta }(\omega )d\omega ,$ we obtain the partial EPI
constants. Values $\Delta _{\alpha }\left( \omega _{n}\right) $ enter to the
expression for the superconducting density of states%
\begin{equation}
N(\omega )=\sum_{\alpha }N_{\alpha }(0)\Real\left\{ \left. \frac{\omega _{n}%
}{\sqrt{\omega _{n}^{2}+\Delta _{\alpha }^{2}\left( \omega _{n}\right) }}%
\right\vert _{i\omega _{n}\rightarrow \omega +i\delta }\right\} .
\label{DOS}
\end{equation}

The Eliashberg functions satisfy the following symmetry relations
\begin{equation}
N_{\alpha }(0)\alpha _{\alpha \beta }^{2}(\omega )F_{\alpha \beta }(\omega
)=N_{\beta }(0)\alpha _{\beta \alpha }^{2}(\omega )F_{\beta \alpha }(\omega
).  \label{sym}
\end{equation}

For $T=T_{c}$ we have (we also neglect the Coulomb pseudopotential)%
\begin{equation*}
\Delta _{\alpha }\left( \omega _{n}\right) Z_{\alpha }\left( \omega
_{n}\right) =\pi T\sum_{\beta }\sum_{\omega _{m}}\frac{\lambda _{\alpha
\beta }\Delta _{\beta }\left( \omega _{m}\right) }{\left\vert \omega
_{m}\right\vert },
\end{equation*}

\begin{equation*}
Z_{\alpha }\left( \omega _{n}\right) =1+\frac{\pi T}{\omega _{n}}%
\sum_{\gamma }\sum_{\omega _{m}}\lambda _{\alpha \gamma }sign\omega _{m},
\end{equation*}%
or%
\begin{eqnarray*}
&&\Delta _{\alpha }\left( \omega _{n}\right) \left[ 1+\frac{\pi T}{\omega
_{n}}\sum_{\gamma }\sum_{\omega _{m}}\lambda _{\alpha \gamma }sign\omega _{m}%
\right] \\
&=&\pi T\sum_{\beta }\sum_{\omega _{m}}\frac{\lambda _{\alpha \beta }\Delta
_{\beta }\left( \omega _{m}\right) }{\left\vert \omega _{m}\right\vert }.
\end{eqnarray*}

Finally

\begin{equation}
\tilde{\Delta}_{\alpha }\left( n\right) \rho (T_{c})=\sum_{\beta
}\sum_{n^{\prime }\geq 1}B_{\alpha \beta }(n,n^{\prime })\tilde{\Delta}%
_{\beta }\left( n^{\prime }\right) ,  \label{eqTc}
\end{equation}%
where for $n,n^{\prime }\geq 1$ the matrix $B_{\alpha \beta }(n,n^{\prime })$
has a form (we have used the symmetry of the gap function $\Delta _{\alpha
}\left( \omega _{n}\right) =\Delta _{\alpha }\left( -\omega _{n}\right) $)
\begin{eqnarray}
B_{\alpha \beta }(n,n^{\prime }) &=&\frac{\lambda _{\alpha \beta
}(n-n^{\prime })+\lambda _{\alpha \beta }(n+n^{\prime }+1)}{\sqrt{%
(2n-1)(2n^{\prime }-1)}}  \notag \\
&&-\delta _{\alpha \beta }\delta _{nn^{\prime }}S_{\beta }(n),
\label{matrB1}
\end{eqnarray}%
where
\begin{equation}
S_{\beta }(n)=\frac{1}{2n-1}\sum_{\gamma }\left[ \lambda _{\beta \gamma
}(0)+2\sum_{m=1}^{n-1}\lambda _{\beta \gamma }(m)\right] ,  \label{S}
\end{equation}%
and $\tilde{\Delta}_{\alpha }\left( n\right) =\Delta _{\alpha }\left(
n\right) /\sqrt{2n-1}$. Here for the simplest Einstein spectrum with the
frequency $\Omega$ $\lambda_{\alpha \beta }(m)=\lambda _{\alpha \beta
}\Omega ^{2}/(\Omega ^{2}+4\pi ^{2}T_{c}^{2}m^{2})$. The value of $T_{c}$ is
determined by the equation
\begin{equation}
\rho (T_{c})=1.  \label{Tc}
\end{equation}

\section{\protect\bigskip Strong coupling}

The simplest way to estimate $\rho (T_{c})$ for superstrong coupling is to
put in Eq.\ref{eqTc} $n=n^{\prime }=1$. In this case we have%
\begin{equation}
\tilde{\Delta}_{\alpha }\left[ \rho (T_{c})+\sum_{\gamma }\lambda _{\alpha
\gamma }\right] =\sum_{\beta }\lambda _{\alpha \beta }\left[ \Omega
^{2}/\left( 2\pi T_{c}\right) ^{2}+1\right] \tilde{\Delta}_{\beta }.
\label{2x2}
\end{equation}%
In the isotropic system $\lambda _{\alpha \beta }=\lambda \delta _{\alpha
\beta }$ the last two terms in the both sides of the equation cancel each
other and we have a standard expression for superstrong coupling (see Refs.%
\cite{AllenMitr},\cite{carbotte})

\begin{equation}
T_{c,iso}=\frac{\Omega }{2\pi }\sqrt{\lambda }  \label{iso}
\end{equation}

For the nondiagonal matrix $\lambda _{\alpha \beta }$ this cancellation does
not occur and the large $\lambda _{\alpha \beta }$ terms play a role of
pair-breaking (see Appendix A).

Let us consider, for the sake of simplicity, the two-band system. The
solution of Eq. (\ref{Tc}) has a form

\begin{equation*}
T_{c,2b}=\frac{\Omega }{2\pi }\sqrt{\frac{A+\newline
\sqrt{B^{2}+4CD}}{2C}}
\end{equation*}%
with the eigenvector%
\begin{equation}
\left\{ \tilde{\Delta}_{1},\tilde{\Delta}_{2}\right\} =\left\{ \frac{%
A^{\prime }+\newline
\sqrt{B^{2}+4CD}}{2\lambda _{21}C},1\right\},  \label{eigen}
\end{equation}
where $A=\lambda _{21}\lambda _{11}+\lambda _{11}+2\lambda _{12}\lambda
_{21}+\lambda _{12}\lambda _{22}+\lambda _{22},$ $A^{\prime }=\lambda
_{11}(1+\lambda _{21})-\lambda _{22}(1+\lambda _{12}),$ $B=\lambda $$%
_{21}\lambda $$_{11}+\lambda $$_{11}+2\lambda $$_{12}\lambda $$_{21}+\lambda
$$_{12}\lambda $$_{22}+\lambda $$_{22},$ $C=1+\lambda $$_{12}+\lambda $$%
_{21},$and $D=\lambda $$_{12}\lambda $$_{21}-\lambda $$_{11}\lambda $$_{22}$,

In this case in the order of $O(1/\lambda )$ (we suppose $\lambda _{11}\sim
\lambda _{12}\sim \lambda _{21}\sim \lambda _{22}\sim \lambda \gg 1$) $%
\tilde{\Delta}_{1}=\tilde{\Delta}_{2}$ and

\begin{equation}
T_{c,2b}\simeq \frac{\Omega }{2\pi }\sqrt{\left\langle \lambda \right\rangle
},  \label{2b}
\end{equation}%
where $\left\langle \lambda \right\rangle $ means averaging over both bands

\begin{equation*}
\left\langle \lambda \right\rangle =\frac{\left( \lambda _{11}+\lambda
_{12}\right) N_{1}(0)+\left( \lambda _{22}+\lambda _{21}\right) N_{2}(0)}{%
N_{1}(0)+N_{2}(0)},
\end{equation*}%
and $N_{\alpha }(0)$ are the partial densities of states. For the general
(non-Einstein ) spectrum we have $\left\langle \lambda \right\rangle \Omega
^{2}\Longrightarrow \left\langle \left[ \lambda \Omega ^{2}\right] _{\alpha
\beta }\right\rangle =\left\langle M_{\alpha \beta }(1)\right\rangle
=\left\langle 2\int_{0}^{\infty }d\omega \omega \left[ \alpha ^{2}(\omega
)F(\omega )\right] _{\alpha \beta }\right\rangle $. This means that strong
coupling leads to washing out effects of anisotropy.

Similar statements were made in Refs. \cite{JC} and \cite{Comb} where the
authors have considered the momentum dependent interaction. In the former
paper the separable interaction similar to Eq.\ref{sep} $\left[ \alpha
^{2}(\omega )F(\omega )\right] _{\mathbf{pp}^{\prime }}=\alpha ^{2}(\omega
)F(\omega )g(\mathbf{p})g(\mathbf{p}^{\prime })$ with $\left\langle g(%
\mathbf{p})\right\rangle =1$. They got the result that the the expression
for $T_{c}$ in the superstrong limit reduces to the isotropic one, while the
"pairing potential" is proportional to $g(\mathbf{p})$. This contradicts to
more general statement in the latter article, where the \textit{positive }%
(attractive) interactions $\alpha ^{2}(\mathbf{k},\mathbf{k}^{\prime
},\omega )F(\mathbf{k},\mathbf{k}^{\prime },\omega )$ for all $\mathbf{k},%
\mathbf{k}^{\prime }$ was investigated and it was shown that the gap
function becomes the $\mathbf{k}$ -independent that leads to the isotropic
expression for $T_{c}.$ The detail inspection the situation in Ref. \cite{JC}
also shows that the real order parameter which enters to the density of
states (see Eq.(\ref{DOS})) is isotropic for large $\lambda $.

We have investigated numerically the evolution of $T_{c}$ and eigenvectors $
\Delta _{\alpha }$ as functions of the coupling strength $\lambda _{\alpha
\beta }$ for the model matrix of the Eliashberg functions

\begin{equation}
\alpha _{\alpha \beta }^{2}(\omega )F_{\alpha \beta }(\omega )=\alpha
^{2}(\omega )F(\omega )\left(
\begin{array}{cc}
1 & 1/5 \\
1/10 & 0%
\end{array}%
\right) .  \label{ml}
\end{equation}

We suppose, for simplicity, 2$N_{1}(0)=N_{2}(0)$ and $\lambda _{22}=0$ (i.e.
no intrinsic superconductivity in the second band). The average EPI constant
$\left\langle \lambda \right\rangle $ is equal to $0.467\lambda $. Results
for $T_{c}$ are presented in Fig.1. We see that for weak and intermediate
coupling there is an enhancement of $T_{c}$ due to anisotropic effects in
comparison with averaged value. \ For small EPI the result coinside with the
weak-coupling expression for $\lambda _{eff}=\lambda _{\max \text{ e.v.}%
}=1.02\lambda $, where $\lambda _{\max \text{ e.v.}}$ is a maximal
eigenvalue of the matrix (\ref{ml}).\ This enhancement, however, vanishes
for large values of $\lambda $ when phonons lead to isotropization of the
superconducting order parameter.

\begin{figure}
\begin{center}

\includegraphics[width=9cm,clip]{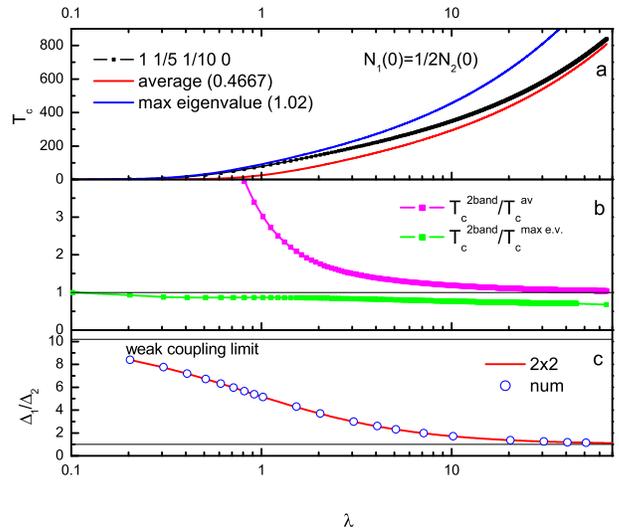}

\end{center}
\caption{(color online) Critical temperature (a) and the gap ratio (c) in the two-band case as a function
of intraband coupling constant $\lambda$ in the first band. The panel (b) shows that $T_c$
in the strongly coupled regime is determined by the average coupling constant. The panel (c)
shows the ratio of the order parameters in the two bands. The gap function becomes isotropic
in the strongly coupled regime. The numerically calculated ratio $\Delta_1/\Delta_2$ is
very accurately described by the expression Eq.(\ref{eigen}) in a broad range of $\lambda$.
}
\label{fig:f1}
\end{figure}

\bigskip We have to note that the result (\ref{2b}) is obtained under the
condition of nonvanishing $\left\langle \lambda \right\rangle $ and in the
Born approximation \cite{dolg1999} for the spin-independent interaction.

Recently the model for the system with strong coupling anisotropic
interaction was considered in Ref. \cite{Umm}. It was supposed that the
difference between the interaction in the quasiparticle channel $%
\left\langle \alpha ^{2}(\mathbf{k},\mathbf{k}^{\prime },\omega )F(\mathbf{k}%
,\mathbf{k}^{\prime },\omega )\right\rangle _{FS}$ and in the Cooper channel
$\left\langle \Delta (\mathbf{k})\alpha ^{2}(\mathbf{k},\mathbf{k}^{\prime
},\omega )F(\mathbf{k},\mathbf{k}^{\prime },\omega )\Delta (\mathbf{k}%
^{\prime })\right\rangle _{FS}/\left\langle \Delta ^{2}(\mathbf{k}%
)\right\rangle _{FS}$ is independent on the coupling strength. Above
analyzes (as well as \cite{Comb,JC}) shows that this difference vanishes for
strong coupling. This removes unphysical results for $T_{c}$ obtained in
this limit in the mentioned paper.

In Appendix B the sensitivity of $T_{c}$ to different phonon modes is
considered by calculating the variational derivatives. It is shown that the
negative (divergent at small frequencies) contribution to the nondiagonal
variational derivative of $T_{c}$ vanishes in the strongly coupled regime.

\section{\protect\bigskip Conclusions}

We have shown that strong coupling effects in the multiband superconductors
lead to the appearance of the strong damping which results from
pair-breaking due to interband coupling.

For systems with the attractive interaction this effect leads to averaging
of order parameters in different bands. As a result asymptotic behavior of $%
T_{c}$ is described by the well known single-band expression $T_{c}\propto
\sqrt{\left\langle \lambda \Omega ^{2}\right\rangle }=\sqrt{%
2\int_{0}^{\infty }d\omega \omega \left\langle \alpha ^{2}(\omega )F(\omega
)\right\rangle }.$ This means that the upper bound on $T_{c}$ in the
superstrong coupling regime is determined by the averaged coupling constant,
while the higher upper bound corresponding to the maximal eigenvalue of the
matrix $\left[ \lambda \Omega ^{2}\right] _{\alpha \beta }$ is never reached.

\textbf{Acknowledgements}. The authors acknowledge many stimulating
discussions with I.I. Mazin. The work is partically supported by NanoNed
program grant TCS7029.

\section{Appendix A}

We extend the results of Ref. \cite{VIK} for effects of low frequency
intermediate boson modes ( $\Omega \lesssim 2\pi T_{c}$) on the critical
temperature of the multi-band superconductors. On the real frequency axis
the equations for the complex order parameter $\Delta _{\alpha }(\omega )$
and the renormalization function $Z_{\alpha }(\omega )$ have forms ( we
neglect the Coulomb contribution)

\begin{equation}
Z_{\alpha }(\omega )\Delta _{\alpha }(\omega )=\sum_{\beta }\int_{-\infty
}^{\infty }dzK_{\alpha \beta }(z^{\prime },\omega )\Real\frac{\Delta _{\beta
}(z^{\prime })}{z^{\prime }},  \tag{A1}  \label{Ad}
\end{equation}

\begin{equation}
(1-Z_{\alpha }(\omega ))\omega =-\sum_{\beta }\int_{-\infty }^{\infty
}dzK_{\alpha \beta }(z^{\prime },\omega ),  \tag{A2}  \label{AZ}
\end{equation}
where $K_{\alpha \beta }(z^{\prime },\omega )$ is a kernel of the
interelectron interaction via intermediate bosons with the spectral function
$\alpha _{\alpha \beta }^{2}(\Omega )F_{\alpha \beta }(\Omega )$%
\begin{eqnarray*}
K_{\alpha \beta }(z^{\prime },\omega ) &=&\frac{1}{2}\int_{0}^{\infty
}d\Omega \alpha _{\alpha \beta }^{2}(\Omega )F_{\alpha \beta }(\Omega )\times
\label{AK} \\
&&\left[ \frac{\tanh \frac{z^{\prime }}{2T_{c}}+\coth \frac{\Omega }{2T_{c}}%
}{z^{\prime }+\Omega -\omega -i\delta }-\left\{ \Omega \rightarrow -\Omega
\right\} \right] .  \notag
\end{eqnarray*}

Now let us separate the functions $\alpha _{\alpha \beta }^{2}(\Omega
)F_{\alpha \beta }(\Omega )$ on to low energy part $\left( \alpha _{\alpha
\beta }^{2}(\Omega )F_{\alpha \beta }(\Omega )\right) ^{<}$ and the
high-energy one

\begin{eqnarray*}
\alpha _{\alpha \beta }^{2}(\Omega )F_{\alpha \beta }(\Omega ) &=&\left(
\alpha _{\alpha \beta }^{2}(\Omega )F_{\alpha \beta }(\Omega )\right)
^{<}\Theta (2\pi T_{c}-\Omega )+ \\
&&\left( \alpha _{\alpha \beta }^{2}(\Omega )F_{\alpha \beta }(\Omega
)\right) ^{>}\Theta (\Omega -2\pi T_{c}).
\end{eqnarray*}%
The same procedure can be done for the kernel (\ref{AK})

\begin{equation}
K_{\alpha \beta }(z^{\prime },\omega )=K_{\alpha \beta }^{<}(z^{\prime
},\omega )+K_{\alpha \beta }^{>}(z^{\prime },\omega ).  \tag{A3}  \label{Klh}
\end{equation}%
In the first term in the right hand side of Eq.(\ref{Klh}) we can neglect
the frequency $\Omega $ in the denominator. In this case

\begin{equation}
K_{\alpha \beta }^{<}(z^{\prime },\omega )=\frac{\Gamma _{\alpha \beta }^{<}%
}{\pi }\frac{1}{z^{\prime }-\omega -i\delta },  \tag{A4}  \label{Kl}
\end{equation}%
where%
\begin{equation*}
\Gamma _{\alpha \beta }^{<} =\pi \int_{0}^{\infty }d\Omega \left( \alpha
_{\alpha \beta }^{2}(\Omega )F_{\alpha \beta }(\Omega )\right) ^{<}\coth
(\Omega /2T_{c}) \\
\simeq 2\pi \lambda _{\alpha \beta }^{<}T_{c}  \tag{A5}  \label{gaml}
\end{equation*}%
is the matrix of the electron scattering on the low-energy excitations. Now
we use the dispersion relation for the order parameter $\Delta _{\beta
}(\omega )$%
\begin{equation}
i\frac{\Delta _{\beta }(\omega )}{\omega }=-\frac{1}{\pi }\int_{-\infty
}^{\infty }\frac{dz^{\prime }}{\omega -z^{\prime }+i\delta }\Real\frac{%
\Delta _{\beta }(z^{\prime })}{z^{\prime }},  \tag{A6}  \label{disp}
\end{equation}%
which is a consequence of the dispersion relation for the electron Green
function in the Nambu representation. Combining expressions (\ref{Ad},\ref%
{AZ}) with (\ref{Klh}-\ref{disp}) we get
\begin{eqnarray*}
&&\Delta _{\alpha }(\omega )\left[ 1+\sum_{\beta }\frac{i\Gamma _{\alpha
\beta }^{<}}{\omega }+\sum_{\beta }\int_{-\infty }^{\infty }dzK_{\alpha
\beta }^{>}(z^{\prime },\omega )\right]  \label{Eq} \\
&=&\sum_{\beta }\frac{i\Gamma _{\alpha \beta }^{<}}{\omega }\Delta _{\beta
}(\omega )+\sum_{\beta }\int_{-\infty }^{\infty }dzK_{\alpha \beta
}^{>}(z^{\prime },\omega )\Real\frac{\Delta _{\beta }(z^{\prime })}{%
z^{\prime }}
\end{eqnarray*}%
We see that the low frequency excitations play a role of intraband and
interband static impurities. Intraband $\Gamma _{\alpha \alpha }$ ones drop
out from the Eq. (\ref{Eq}) (so called Anderson's theorem). It is
interesting to note that the famous cancellation of the largest terms
proportional to $\lambda ^{<}$ (see e.g., \cite{AllenMitr}) comes not from
the strong renormalization of the quasiparticle energy ($\Real Z$), but from
the damping $i\Gamma ^{<}$.

\section{Appendix B}

In Ref.\cite{mitr} the sensitivity of $T_{c}$ to different phonon modes was
considered by calculating the variational derivatives $\delta T_{c}/\delta %
\left[ \alpha ^{2}(\Omega )F(\Omega )\right] _{\alpha \beta }$. For the
diagonal elements ($\alpha =\beta $) the result for small $\Omega $ ($\Omega
\ll 2\pi T_{c}$) coincides with the one obtained by Bergmann and Rainer \cite%
{BR} $\delta T_{c}/\delta \left[ \alpha ^{2}(\Omega )F(\Omega )\right]
\thicksim \Omega $ for the isotropic single band system. This corresponds to
the enhancement of $T_{c}$ by adding low frequency phonons (bosons).

In the multiband case the interband derivative has the following form

\begin{equation}
\frac{\delta T_{c}}{\delta \left[ \alpha ^{2}(\Omega )F(\Omega )\right]
_{\alpha \neq \beta }} \thicksim  \notag \\
\end{equation}
\begin{equation}
\frac{N_{\alpha }(0)}{\Omega }\sum_{n\geq 1}\frac{\Delta _{\alpha }\left(
\omega _{n}\right) \left[ \Delta _{\beta }\left( \omega _{n}\right) -\Delta
_{\alpha }\left( \omega _{n}\right) \right] }{\omega _{n}^{2}} \text{\ } +  \text{\ } O(\Omega ),
\notag
\end{equation}
and $\delta T_{c}/\delta \left[ \alpha ^{2}(\Omega )F(\Omega )\right] _{12}$
and $\delta T_{c}/\delta \left[ \alpha ^{2}(\Omega )F(\Omega )\right] _{21}$
have different signs. This contradicts to the symmetry relation (\ref{sym}).
If we change the function $\left[ \alpha ^{2}(\Omega )F(\Omega )\right]
_{12} $ the counterpart $\left[ \alpha ^{2}(\Omega )F(\Omega )\right] _{21}$
has to be changed automatically. Only the symmetrized \textit{off-diagonal}
combination $\delta T_{c}/\delta \alpha ^{2}(\Omega )F(\Omega )_{\text{o.d.}}
$ has physical meaning. Here
\begin{equation}
\alpha ^{2}(\Omega )F(\Omega )_{\text{o.d.}}=  \notag
\end{equation}
\begin{equation}
\frac{N_{1}(0)\left[ \alpha ^{2}(\Omega )F(\Omega )\right] _{12}+N_{2}(0)%
\left[ \alpha ^{2}(\Omega )F(\Omega )\right] _{21}}{N_{1}(0)+N_{2}(0)}.
\notag
\end{equation}
As a result, we obtain
\begin{equation}
\frac{\delta T_{c}}{\delta \alpha ^{2}(\Omega )F(\Omega )_{\text{o.d.}}}
\thicksim -\frac{N_{\alpha }(0)}{\Omega }\sum_{n\geq 1}\frac{\left[ \Delta
_{2}\left( \omega _{n}\right) -\Delta _{1}\left( \omega _{n}\right) \right]
^{2}}{\omega _{n}^{2}}  \notag
\end{equation}
\begin{equation}
+ \text{ \ } O(\Omega ).  \tag{B1}  \label{B1}
\end{equation}

In contrast to the single band case (see \cite{carbotte}), the off-diagonal
derivative has different behavior in the weak-coupling and strong-coupling
regimes. For the former case one can suppose $\Delta _{\alpha }\left( \omega
_{n}\right) =\Delta _{\alpha }\left( \Theta -\left\vert \omega
_{n}\right\vert \right) $ and $\delta T_{c}/\delta \alpha ^{2}(\Omega
)F(\Omega )_{\text{o.d.}}\thicksim -\frac{\left( \Delta _{2}-\Delta
_{1}\right) ^{2}}{\Omega }.$ This means that the addition of nondiagonal
interaction with low frequency phonons leads to strong suppression of
the critical temperature in weak-coupling anisotropic superconductors.
This result was obtained in Ref. \cite{DC} for the anisotropic separable
interaction. In the strong coupling limit, as it was shown above, $\Delta
_{1}\Longrightarrow \Delta _{2}$, then the first term in (\ref{B1}) vanishes
and $\delta T_{c}/\delta \alpha ^{2}(\Omega )F(\Omega )_{\text{o.d.}%
}\thicksim \Omega >0$, similar to the intraband contribution. This result
can be directly obtained from the expression (\ref{2b}).

\end{document}